\documentclass[traditabstract]{aa} 
\usepackage{graphicx}
\usepackage{txfonts}
\begin{document}
   \title{Helium star evolutionary channel to super-Chandrasekhar mass type Ia supernovae}

   \author{W. -M. Liu$^{1}$, W. -C. Chen$^{1,2,3}$, B. Wang$^{4}$, and Z. W. Han$^{4,5}$          }

   \institute{$^1$ Department of Physics, Shangqiu Normal University,
Shangqiu 476000, China;
\\
 $^2$ School of Physics and State Key Laboratory
of Nuclear Physics and Technology, Peking University, Beijing
100871, China\\
$^3$ Key Laboratory of Modern Astronomy and Astrophysics (Nanjing
University), Ministry of Education, Nanjing 210093,
China\\
$^4$ National Astronomical Observatories/Yunnan Observatory, the
Chinese Academy of Sciences, Kunming 650011,
China\\
$^5$ Key Laboratory for the Structure and Evolution of Celestial
Objects, the Chinese Academy of Sciences, Kunming 650011, China\\
\email{chenwc@nju.edu.cn} }

   \date{}


  \abstract
{Recent discovery of several overluminous type Ia supernovae (SNe
Ia) indicates that the explosive masses of white dwarfs may
significantly exceed the canonical Chandrasekhar mass limit. Rapid
differential rotation may support these massive white dwarfs.
Based on the single-degenerate scenario, and assuming that the
white dwarfs would differentially rotate when the accretion rate
$\dot{M}>3\times 10^{-7}M_{\odot}\rm yr^{-1}$, employing
Eggleton's stellar evolution code we have performed the numerical
calculations for $\sim$ 1000 binary systems consisting of a He
star and a CO white dwarf (WD). We present the initial parameters
in the orbital period - helium star mass plane (for WD masses of
$1.0~M_{\odot}$ and $1.2~M_{\odot}$, respectively), which lead to
super-Chandrasekhar mass SNe Ia. Our results indicate that, for an
initial massive WD of $1.2~M_{\odot}$, a large number of SNe Ia
may result from super-Chandrasekhar mass WDs, and the highest mass
of the WD at the moment of SNe Ia explosion is 1.81 $M_\odot$, but
very massive ($>1.85M_{\odot}$) WDs cannot be formed. However,
when the initial mass of WDs is $1.0~M_{\odot}$, the explosive
masses of SNe Ia are nearly uniform, which is consistent with the
rareness of super-Chandrasekhar mass SNe Ia in observations.
   }


     \titlerunning{He star evolutionary channel to super-chandrasekhar mass SNe Ia}
\authorrunning{Liu et al.}

   \maketitle


\section{Introduction}

Type Ia supernovae (SNe Ia) are generally believed to be the
thermonuclear explosions of carbon-oxygen white dwarfs (CO WDs) in
binaries (for a review see \cite{nom97}). Because of an uniform
luminosity, SNe Ia are used as the standard candlelight to
determine the cosmological distances, and estimate the
cosmological parameters $\Omega$ and $\Lambda$ (e.g.
\cite{ries98,perl99}). However, some key issues including the
properties of their progenitors and the physical mechanisms of the
explosion are still poorly known by astrophysicists
(\cite{hill00,rop05,wang08,pod08}).

At present, there exist two progenitor models of SNe Ia, i. e.
single-degenerate model (\cite{whel73,nom82}) and
double-degenerate model (\cite{iben84,webb84}). Over the past
decade, many groups have widely investigated the single-degenerate
model, in which the CO WD accretes H/He-rich material from a
non-degenerate companion star. In this scenario, the donor star of
WD may be a main-sequence star, a subgiant star or a red-giant
star (see Hachisu et al. 1996; Li \& van den Heuvel 1997; Hachisu
et al. 1999a,b; Langer et al. 2000; Han \& Podsiadlowski 2004,
2006, Chen \& Li 2007, Han 2008, Chen \& Li 2009, Meng et al.
2009, L\"{u} et al. 2009, Wang et al. 2010a, Meng \& Yang
2010a,b).

\cite{limo91} have firstly studied the evolution of the binary
consisting of a CO WD and a He star, in which the CO WD accretes
He matter from the He star and grows its mass to the canonical
Chandrasekhar limit of $1.4M_{\odot}$. Yoon \& Langer (2003) also
found that CO WD + He star systems can form a reliable channel
producing SNe Ia. Recently, \cite{wang09a} systematically explored
the evolution of 2600 close WD binaries including a He MS star or
He subgiant, and obtained the parameter spaces for the progenitor
of SNe Ia. Using a detailed binary population synthesis approach,
Wang et al. (2009b) suggested that this channel can contribute a
SN Ia birthrate of $3\times 10^{-4}~\rm yr^{-1}$ in the Galaxy,
and can result in the short delay times ($\la100\rm Myr$) between
the formation of the progenitor systems and the explosions.

Recently, overluminous SN Ia 2003fg has been detected
(\cite{asti06}), and its explosive mass was estimated to be $\sim
2.1~M_{\odot}$ (\cite{how06}) \footnote{Hillebrandt et al. (2007)
suggested that a lopsided explosion due to the off-center ignition
of nuclear burning in a WD with Chandrasekhar mass may be
responsible for the high brightness of 2003fg.}. Later, several
possible overluminous SNe Ia 2006gz, 2007if, and 2009dc were also
discovered (\cite{hick07,yuan07,tana10,scal10,silv10}). The
differential rotation may support these massive WDs
(\cite{yoon04a,yoon04b,yoon05}), which may also result from the
merger of two massive WDs (\cite{tutu94,how01,pier03}). Based on
the single degenerate model and the assumption that the WD
differentially rotate, Chen \& Li (2009) have explored the
evolution of close binaries consisting of a WD and a H main
sequence star. Their results indicate that, for an initial massive
WD with $1.2~M_{\odot}$, the maximum explosive mass of the WDs is
$1.76~M_{\odot}$.

It is an interesting issue to explore the progenitors of
overluminous SNe Ia based on the single-degenerate model. In this
paper, we attempt to explore whether the WD + He star systems can
produce overluminous SNe Ia. To obtain the distribution of the
initial He star mass and the initial orbital period of the
progenitor binaries, we have calculated the evolution of 1000 WD
binaries including a He star and a rotating CO WD. In section 2,
we give a detailed description of input physics including the
formation of WD + He star systems and the binary evolution code.
In section 3 we present the calculated results. Finally, a brief
discussion and summary are presented.
\section{Input physics}
\subsection{Formation of WD + He star systems}
In young stellar populations with relatively late star formation,
there exist intermediate-mass binary systems, which can evolve
into CO WD + He star systems. For a large range of initial
parameters, close binaries can evolve into a system consisting of
a more massive WD and a less massive He star
(\cite{torn86,iben87a,iben87b}). The mass transfer rate in such
systems is only $\sim 3\times10^{-8}~M_{\odot}\rm yr^{-1}$, which
may result in an explosion of SN magnitude after the WD
accumulates $\sim 0.15~M_{\odot}$ of helium (\cite{iben91}).
However, Iben \& Tutukov (1994) proposed that WD + massive He star
systems can be formed via binary evolution, and the He shell
burning on the surface of the CO WD could be in a quasi-stationary
after the massive He star fills its Roche lobe.

Using the rapid binary star evolution (BSE) code (Hurley et al.
2000, 2002), we calculated the formation process of a massive CO
WD + He star system. We consider a primordial binary consisting of
a primary (star 1) of $M_{1}=7.5~M_{\odot}$ and a star 2 of
$M_{2}=4.0~M_{\odot}$ ($Z=0.02$), and with an orbital separation
of $a=50.93~R_{\odot}$. Owing to the nuclear evolution, when
\textbf{$t=42.42~\rm Myr$} the primary firstly fills its Roche
lobe, and transfers its material to the star 2. Because of a
relatively high initial mass ratio, the mass transfer occurs on a
short thermal timescale. When $t=42.52~\rm Myr$,
$M_{1}=1.482~M_{\odot}$, $M_{2}=10.001~M_{\odot}$, and the first
Roche lobe overflow ceases. When $t=52.20~\rm Myr$, the primary
that evolves into a He star fills its Roche lobe again, and then
evolves into a CO WD of $M_{1}=1.103~M_{\odot}$ after the second
Roche lobe overflow ceases. At the same time, the star 2 locates
Hertzsprung gap, and $M_{2}=10.194~M_{\odot}$. When $t=64.33~\rm
Myr$, the star 2 continues to evolve and fill its Roche lobe. Due
to the ultra-high mass ratio, a common envelope is possibly formed
quickly because of dynamically unstable mass transfer. After the
common envelope is ejected, a close binary system consisting of a
CO WD ($M_{1}=1.103~M_{\odot}$) and a  He star
($M_{2}=2.247~M_{\odot}$) is then formed. This formation channel
is known as the He star channel(for details see Wang et al.
2009b), which allows stable Roche lobe overflow to produce massive
WD + He star system (rather lead to dynamical mass transfer and a
common envelope phase).

In addition, other two channels such as early asymptotic giant
branch channel and thermal pulsing asymptotic giant branch channel
can also produce CO WD + He star systems (\cite{wang09b}).
Employing Hurley's rapid BSE code, Wang et al. ( 2009b )
investigated the initial parameters distribution  of WD + He star
systems. Their simulated results show that, the initial mass of
the CO WDs is in the range of $0.85 - 1.2~ M_\odot$
\footnote{Rapid binary evolution code presents a rough statistical
result, while standard stellar evolutionary theory suggested that
$1.1~M_{\odot}$ is the mass maximum of the CO core. Assuming that
the envelope of AGB star is lost when the binding energy of the
envelope is equal to zero, Han et al (1994) and Meng et al (2008)
found that the maximum mass of CO WD is $1.1 M_{\odot}$ and $1.05
M_{\odot}$ ($Z=0.02$), respectively. If a CO core with $1.2
M_{\odot}$ is produced by stellar evolution, a C burning should
occur, and result in the formation of ONeMg WD. Therefore, it is
very difficult to form a $1.2 M_{\odot}$ CO WD from a stellar
evolutionary channel. Certainly, the maximum mass of the CO WD is
related to the definition of the core, and the necessary condition
ejecting the envelope of an AGB star. In addition, the lifting
effect caused by rotation may help in producing larger CO core
masses. We expect further detailed study to confirm if the most
massive CO WD can exceed $\sim1.1 M_{\odot}$ as a result of binary
interaction. Similar to Hachisu et al. (1996) and Li \& van den
Heuvel (1997), in this work we still take a maximum initial mass
of $1.2 M_{\odot}$ for WDs.}, and the initial mass of the He stars
is in the range of $1.0 - 2.8 ~M_\odot$, which depend on the
common envelope parameters $\alpha_{\rm CE}\lambda$. Figure 1
presents the distribution of the initial WD + He star systems in
$M_{\rm He}-M_{\rm WD}$ diagram, in which we have followed the
evolution of $4\times10^{7}$ sample binaries from the star
formation to the formation of the WD + He star systems and adopt
$\alpha_{\rm CE}\lambda=0.5$. Based on this work, for input
parameters we take the initial mass of the WDs $M_{\rm WD,i}=1.0,
1.2~M_\odot$, the initial mass of the He stars $M_{\rm He,i}=1.0 -
2.8~M_\odot$, and the initial orbital period $\rm{log}(P_{\rm
orb,i}/d)=-1.4-0$ (see Fig. 3 in Wang et al. 2009b).

\begin{figure}
\centering
\includegraphics[angle=0,width=9.0cm]{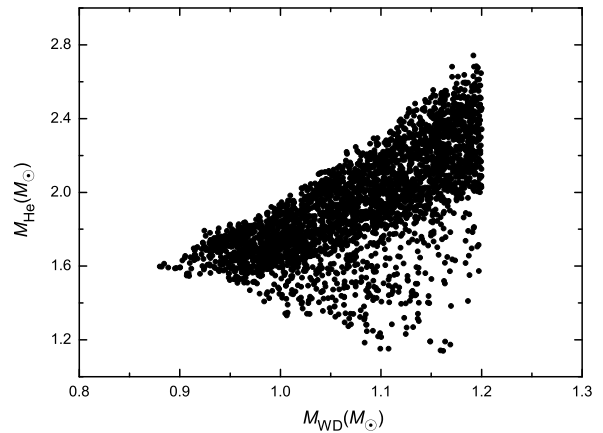}
\caption{Distribution of initial WD + He star systems in $M_{\rm
He}-M_{\rm WD}$ diagram. Basic input parameters for Monte Carlo
simulations correspond to the standard model of Wang et al.
(2009b).} \label{FigVibStab}
\end{figure}
\subsection{Binary evolution code}
Employing stellar evolution code developed by Eggleton (see
Eggleton 1971, 1972, 1973), which has been updated with the latest
input physics (\cite{han94,pol95,pol98}), we study the evolution
of close binaries consisting of a rotating CO WD (of mass $M_{\rm
WD}$) and a He star (of mass $M_{\rm He}$). The stellar OPAL
opacities are come from \cite{roge92}, and \cite{alex94} for a low
temperature. In our calculations, we set the ratio of mixing
length to local pressure scale height to be $2.0$, and the
overshooting parameter of the He stars (with chemical abundance $Y
= 0.98$,  $ Z = 0.02$) is taken to be 0 (\cite{dewi02}).

In a WD + He star system, due to the nuclear evolution or the loss
of orbital angular momentum the He star fills its Roche lobe at He
main sequence or He subgiant stage, and then begins the mass
transfer. The He-rich material from the He star is accreted by the
WD, and is converted into CO heavier elements via thermonuclear
burning onto the surface of the WD. In input physics, we adopt an
optically thick wind scenario (\cite{kato94,hach96}) and the
description for He mass accumulation efficiency onto the surface
of the WD given by Kato \& Hachisu (2004). Firstly, if the mass
transfer rate $|\dot{M}_{\rm He}|$ is above a critical rate
(\cite{nom82})
\begin{equation}
\dot{M}_{\rm {cr}}=7.2\times10^{-6}({M_{\rm
WD}}/{M_{\odot}}-0.6)\quad M_{\odot}\,\rm yr^{-1},
\end{equation}
we take that He burning is steady, and the He-rich material on the
surface of the WD is converted into CO elements at a rate
$\dot{M}_{\rm {cr}}$.  Secondly, if $|\dot{M}_{\rm He}|$ is lower
than $\dot{M}_{\rm {cr}}$ but higher than $\dot{M}_{\rm {st}}$
(Kato \& Hachisu 2004), which is the minimum accretion rate that
the He-shell steadily burn, the He burning is thought to be
steady, and all accreting material is converted into CO elements.
Thirdly, if $|\dot{M}_{\rm He}|$ is lower than $\dot{M}_{\rm
{st}}$ but higher than $\dot{M}_{\rm {low}}=4.0\times
10^{-8}~M_{\odot}\,\rm yr^{-1}$, in which the weak He-shell
flashes occur \footnote{When weak He-shell flashes occur, the mass
accumulation efficiency strongly depend on the chemical
composition or the radius of the WD (Kato \& Hachisu 2004).}, a
part of the envelope material of the WD is assumed to be blown off
from the surface of the WD (\cite{woos86}). Finally, if
$|\dot{M}_{\rm He}|$ is lower than $\dot{M}_{\rm {low}}$, the
strong He-shell flashes occur, and no material can be accumulated
onto the WD. Summarizing the above prescriptions, the accumulation
efficiency of the accreting He can be written as (Wang et al.
2009a)
\begin{equation}
\alpha=\left\{
\begin{array}{l}
\frac{\dot{M}_{\rm {cr}}}{|\dot{M}_{\rm He}|},\quad |\dot{M}_{\rm He}|>\dot{M}_{\rm {cr}} ,\\
1,\qquad\dot{M}_{\rm {cr}}\geq{|\dot{M}_{\rm He}|}\geq{\dot{M}_{\rm {st}}} ,\\
\alpha',\quad\dot{M}_{\rm {st}}>{|\dot{M}_{\rm He}|}\geq{\dot{M}_{\rm {low}}} ,\\
0,\qquad|\dot{M}_{\rm He}|<\dot{M}_{\rm {low}},
\end{array}
\right.
\end{equation}
where $\alpha'$ is determined by a linearly interpolated way from
a grid computed by \cite{kato04}.

Actually, the accumulation efficiency mentioned above is limited
to non-rotating WDs, and is inadequate to calculate the mass
increase of rotating WD. Considering the spin-up of the WDs via
accretion, Yoon \& Langer (2004) found that the rotation can
stabilize He-shell burning, and help He-accreting CO WD grow in
mass. Assuming that the CO WDs rigidly rotate, the simulated
results by \cite{Domi06} proposed that massive progenitors result
in higher $^{56}$Ni mass and explosive luminosity, and more
massive WDs at the moment of explosion.

To consider the influence of rotation on the mass accumulation of
WDs, we adopt the following input physics (e.g. \cite{chen09}).
(1) With the mass transfer from the He star, the WDs obtain a
large amount of angular momentum from the accreting material, and
is spun up to a high rotation velocity
(\cite{duri77,ritt85,nara89,lang00}). (2) Considering the lifting
effect in the hydrostatic equilibrium (\cite{Domi96}), we
introduce an effective mass $M_{\rm eff}$ of the WD by taking
account of the centrifugal force. (3) According to different
ranges of the polar angle, we divide the surface of the WD into
three zones as follows : the equatorial zone (EZ,
$\theta=60^{\circ}-120^{\circ}$), the middle zone (MZ,
$\theta=30^{\circ}-60^{\circ} $ and $120^{\circ}-150^{\circ} $),
and the polar zone (PZ, $\theta=0^{\circ}-30^{\circ}$ and
$150^{\circ}-180^{\circ} $). (4) Assuming that each zone accretes
the transferred material at a rate proportional to its area, we
can obtain accretion fraction $f_{\rm i}=0.5$, 0.366, and 0.134
for EZ, MZ, and PZ, respectively. Therefore, the mass growth rate
of the rotating WD is given by \footnote{In this work, we have not
calculated the detailed evolution of the WD, and only consider the
mass accumulation process on the surface.}
\begin{equation}
\dot{M}_{\rm WD}= \sum \alpha_{\rm i}f_{\rm i}|\dot{M}_{\rm He}|,
\end{equation}
where $\alpha_{\rm i}$ is the mass accumulation efficiencies for
different zones on the surface of the WD. The mass loss rate of
the binary system is $\dot{M}=(1-\sum \alpha_{\rm i}f_{\rm
i})\dot{M}_{\rm He}$, which is assumed to be ejected in the
vicinity of the WD in the form of isotropic winds or outflows, and
taking away the specific orbital angular momentum of the WD.

In our calculations, we set the initial surface velocity at the
WD's equator to be $10~ \rm km\,s^{-1}$, and the radius of the WD
changes with $R\propto M_{\rm WD}^{-1/3}$. The criterion that the
WD differentially rotate is a key input physics in this work. We
adopt the conclusion derived by \cite{yoon04b}, in which the WD
should differentially rotate when its accretion rate
$\ga3\times10^{-7}M_\odot$ y$\rm {r}^{-1}$. As a result of
differential rotation, the central carbon ignition of the WD
cannot occur even if its mass exceed canonical Chandrasekhar limit
of $1.4~M_\odot$. Once $M_{\rm WD}\ge1.4~M_\odot$ and $\dot
M<3\times10^{-7}~M_\odot$, we stop the calculation, and assume the
WD to explode as a SN Ia (we use $M_{\rm SN}$ to denote the
explosive mass of the WD) because of no differential rotation to
support the massive WD. Considering the prescriptions above for
the mass accumulation on the surface of the WD in Eggleton¡¯s
stellar evolution code, we have calculated the evolution of WD +
He star systems, and obtained the initial parameters of the WD
binaries that lead to SNe Ia.

\section{Binary evolution results}
An example of the evolutionary sequences of a WD binary (with
$M_{\rm He, i}=1.8~M_\odot$, $M_{\rm WD, i}=1.2~M_\odot$, and
log($P_{\rm orb, i}$/day) = -1.20) are shown in Figures 2 and 3.
We plot the evolution of $\dot {M}_{\rm He}$, ${\dot M}_{\rm WD}$
and $M_{\rm WD}$ varying with time in Figure 2. When the age of
the He star is $1.76\times 10^{6}$ yr, the Roche lobe overflow
occurs, and the WD accretes the material from the He star. In the
earlier phase of the mass transfer, the orbital period decreases
until log($P_{\rm orb, i}$/day $\approx -1.24$, because the
material is transferred from the more massive He star to the less
massive WD, and then increases when the WD mass grows and exceeds
the He star mass. After the mass exchange of $2.75\times 10^{5}$
yr, the mass transfer rate decline to be $3\times 10^{-7}M_{\odot}
\rm yr^{-1}$, and the WD cannot differentially rotate and SN Ia
explosion is triggered. At the moment of the SN explosion, the WD
grows to $1.81~M_\odot$, the mass of the He star is
$0.96~M_\odot$, and log($P_{\rm orb}$/day) = -1.06. During the
stable He shell burning, the mass accretion rate of the WD is
$10^{-7}-10^{-5}~M_\odot\,\rm yr^{-1}$. By $L_{\rm
X}\sim\epsilon_{\rm He}|\dot{M}_{\rm He}|$ ($\epsilon_{\rm
He}=6.0\times10^{17}\rm erg\,g^{-1}$ is the released energy rate
by He burning), we can estimate the X-ray luminosity of the WD +
He star system to be $\sim10^{36}-10^{38}\rm erg\,s^{-1}$. This
estimation is consistent with the observed X-ray luminosity for
supersoft X-ray sources (SSSs, \cite{kaha97}), and confirms the
conclusion that WD + He star systems may appear as SSSs before SN
Ia explosions (\cite{iben94,yoon03}).
\begin{figure}
\centering
\includegraphics[angle=0,width=9cm]{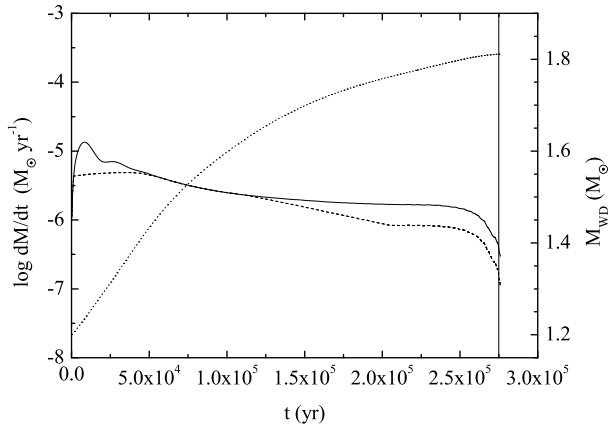}
\caption{Evolution tracks of WD binary with $M_{\rm He, i}=1.8
~M_\odot$, $M_{\rm WD,i}=1.2~M_\odot$, and log($P_{\rm orb,i}/{\rm
day})=-1.2$. The solid, dashed, and dotted curves represent the
evolution of $\dot {M}_{\rm He}$, ${\dot M}_{\rm WD}$, and $M_{\rm
WD}$, respectively. The solid vertical line indicates the position
where the WD is expected to explode as a SN Ia.}
\label{FigVibStab}
\end{figure}

\begin{figure}
\centering
\includegraphics[angle=0,width=9cm]{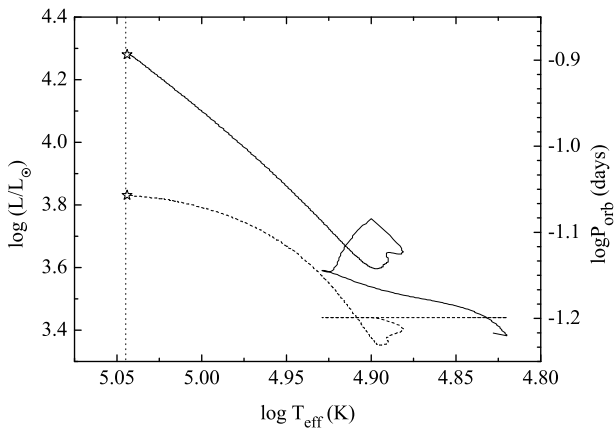}
\caption{Evolutionary track of the He star is shown as a solid
curve, and the evolution of orbital period is shown as a dashed
curve. The vertical dotted line and open stars indicate the place
where the WD explodes as a SN Ia. } \label{FigVibStab}
\end{figure}

To obtain the distribution of the initial He star mass and the
initial orbital period for the progenitors of super-Chandrasekhar
mass SNe Ia, we have calculated the evolutions of about 1000 WD
binaries with different input parameters, which include the
initial mass of the donor star $M_{\rm He, i}$, the initial
orbital period $P_{\rm orb, i}$, and the initial mass of the WD
$M_{\rm WD,i}$ (we take to be $1.2M_\odot$ and $1.0M_\odot$).
Figure 4 shows our calculation results in $M_{\rm He,i}$ -
log$P_{\rm orb, i}$  diagram when $M_{\rm WD,i}=1.2~M_\odot$, and
the regions enclosed by the solid, dashed, and dotted curves
represent the distribution areas of initial WD binaries with
$M_{\rm SN}\geq1.7~M_\odot$, $M_{\rm SN}\geq1.6~M_\odot$, and
$M_{\rm SN}\geq1.4~M_\odot$, respectively. Beyond the boundary
enclosed by the dotted curve, the mass of the WD cannot reach the
Chandrasekhar limit due to either a low mass accumulation
efficiency in the surface of the WD or unstable mass transfer. In
particular, there exist 8 WD binaries with an explosive mass of
$M_{\rm SN}\geq1.8~M_\odot$ (see the star signs in Figure 4),
while the maximum explosive mass is $1.76~M_\odot$ for WD binaries
with a main sequence or sub-giant companion (\cite{chen09}). When
$M_{\rm WD,i}=1.0~M_\odot$, the progenitors distribution of SNe Ia
in the $M_{\rm He, i}-{\rm log}P_{\rm orb, i}$ diagram is plotted
in Figure 5, in which the solid, and dashed curves represent the
region boundaries with $1.5~M_\odot \geq M_{\rm
SN}\geq1.45~M_\odot$, and $M_{\rm SN}\geq1.4~M_\odot$,
respectively. In this case, the mass of the WD cannot grow to
$\geq 1.5~M_\odot$. It is clearly seen that, for a WD with $M_{\rm
WD,i}=1.2~M_\odot$, its maximum accumulated mass is greater than
those of the WD with $M_{\rm WD,i}=1.0~M_\odot$ by more than
$\sim0.1~M_\odot$. This difference mainly originates from the
various mass accumulation efficiency on the surface of the WD. One
can see that the higher the WD mass is, the higher the critical
mass transfer rate is, and hence the higher mass accumulation
efficiency under an approximate equal mass transfer rate (see
Eqs.~1 and 2).

Figures 6 and 7 present the distribution of the donor stars at the
moment of SNe Ia in the H-R diagram when $M_{\rm
WD,i}=1.2~M_{\odot}$, and $M_{\rm WD,i}=1.0~M_{\odot}$,
respectively. They have luminosities in the range of ${\rm
log}(L/L_{\odot})\sim 3.2-4.4$ and effective temperatures in the
range of $10000-150000\rm K$. These properties may be compared
with and testified by future optical observations of SN Ia
remnants.

\begin{figure}
\centering
\includegraphics[angle=0,width=9cm]{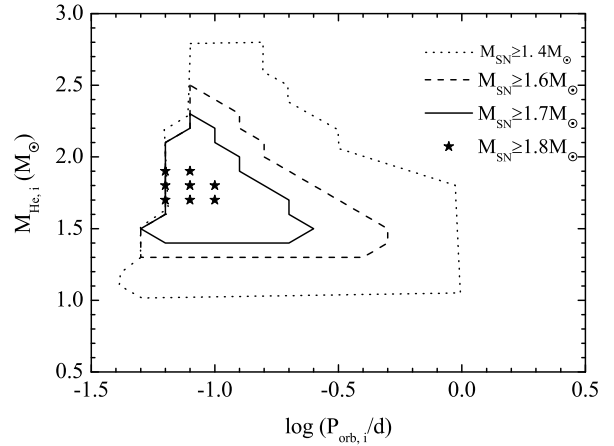}
\caption{Distribution of the initial orbital periods $P_{\rm orb,
i}$ and the initial He star masses $M_{\rm He,i}$ of the
progenitors of super-Chandrasekhar mass SNe Ia when $M_{\rm
WD,i}=1.20~M_{\odot}$. } \label{FigVibStab}
\end{figure}

\section{Discussion and Summary}

Overluminous SNe Ia may be originated from the thermonuclear
explosion of super-Chandrasekhar mass WDs, which may be supported
by rapid differential rotation. Enlightening by the works such as
Wang et al (2009a, 2009b, 2010b) and Chen \& Li (2009), in this
paper we investigate the evolutionary channel of He star + WD with
differential rotation, which may be the progenitors of
super-Chandrasekhar mass SNe Ia like SN Ia 2003fg. Assuming that
the WD differentially rotate when
$\dot{M}>3\times10^{-7}~M_{\odot}\,\rm yr^{-1}$, we have performed
the numerical calculations for $\sim1000$ WD binaries consisting
of a WD and a He star.

\begin{figure}
\centering
\includegraphics[angle=0,width=9.0cm]{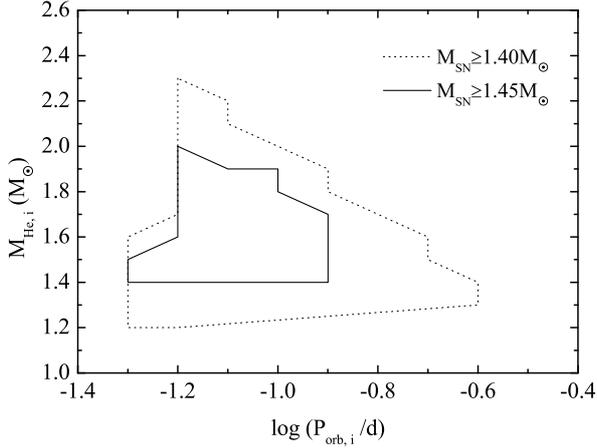}
\caption{Similar to Fig.~4, but for an initial WD mass of
$1.0~M_\odot$. } \label{FigVibStab}
\end{figure}

Our main results can be summarized as follows:
 \begin{enumerate}
\item When the initial mass of the WDs $M_{\rm WD,i} = 1.2~
M_\odot$, the explosive masses of the WDs are in the range of
$1.40- 1.81~ M_\odot$. This implies that a large number of SNe Ia
may result from super-Chandrasekhar mass WDs, while very massive
($M_{\rm SN}>1.85M_{\odot}$) WDs cannot be formed. Same as
 Chen \& Li (2009), our He star evolutionary channel cannot
produce super-Chandrasekhar mass SN Ia with $M_{\rm
SN}\sim2.1M_{\odot}$ like 2003fg (\cite{how06}). It seems that the
merger of two massive WDs may be responsible for this peculiar SN
Ia (\cite{tutu94,how01}).

\item When $M_{\rm WD,i} = 1.0~ M_\odot$, all the explosive masses
of the WDs are less than $1.5~M_\odot$, i.e. the explosive
luminosity of SNe Ia are nearly uniform. These results are
consistent with the rareness of overluminous SNe Ia in
observations.

\item Comparing with the main sequence star, subgiant star, and
red-giant channel, He star channel has higher explosive masses,
less orbital separation, and shorter mass transfer timescale.

\end{enumerate}

The WD + He star systems may be the candidate of SSSs. \cite{di03}
found that some SSSs in spiral galaxies are associated with spiral
arms, which shows that they are young systems with an age of
$\la0.1\rm Gyr$. The WD + He stars channel can explain SNe Ia with
short delay times ($ \la 10^{8}$ yr) (Wang et al. 2009a, 2009b,
\cite{meng10b}), hence this channel may be related to the young
SSSs. In addition,  WD + He star channel may be responsible for
the origin of the hypervelocity stars (\cite{wang09c}).

HD 49798/RXJ 0648.0-418 is an evidence of the existence of rapidly
rotating massive WD + He star system. Based on data from the XMM -
Newton satellite, the masses of the hot subdwarf HD 49798 and the
WD are constrained to be $1.50\pm0.05~ M_\odot$ and $1.28\pm0.05~
M_\odot$, respectively (\cite{mer09}). \cite{wang10c} suggested
that the hot subdwarf HD 49798 and its X-ray pulsating companion
could evolve into an SN Ia in the future . V445 Pup is another
candidate of massive WD + He star system, in which the mass of the
WD is $\ga1.35~M_\odot$, and the mass of the He star is
$\ga0.85~M_\odot$ (\cite{kato08,wou09}).

The formation mechanism of the overluminous SNe Ia is still an
open question. Both the merger of two massive WDs and the
differentially rotating WDs in a close binary may be the
progenitors of overluminous SNe Ia. For the latter, the mass
accumulated efficiencies onto the surface of the WD with
differential rotation need further detailed numerical simulation.
In this paper, we adopt a simple procedure, whereas the actual
process of the thermonuclear burning on the surface of the WDs
should be very complicated. However, the He star evolutionary star
channel may play an important role in the formation of brighter
SNe Ia, which are more frequent in the young stellar population
(\cite{hamu95,hamu96,aubo08}). We expect further detailed
multi-waveband observations for the companions of SNe Ia to
confirm the He star evolutionary channel in the future, though it
is predicted that SNe Ia from this channel may be rare
(\cite{kato08}).

\begin{figure}
\centering
\includegraphics[angle=0,width=9.0cm]{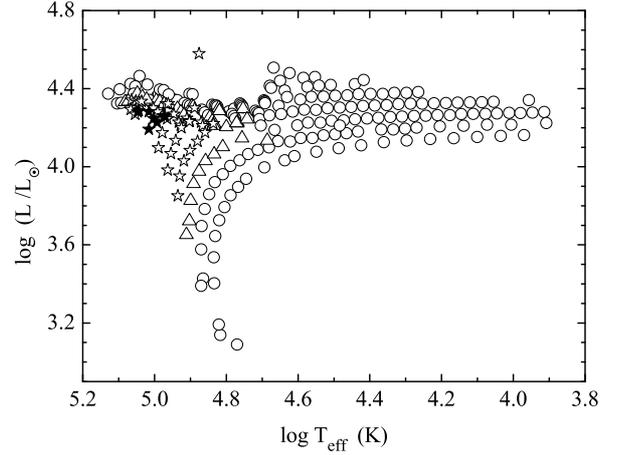}
\caption{Distribution of the donor stars at the momentum of SN Ia
explosions in the H-R diagram when $M_{\rm WD,i}=1.2~M_\odot$. The
open circles, open stars, open triangles, and solid stars denote
systems with $1.4~M_\odot\leq M_{\rm
SN}<1.6~M_\odot,~1.6~M_\odot\leq M_{\rm
SN}<1.7~M_\odot,~1.7~M_\odot\leq M_{\rm SN}<1.8~M_\odot$, and
$M_{\rm SN}\geq1.8~M_\odot$, respectively. } \label{FigVibStab}
\end{figure}

\begin{figure}
\centering
\includegraphics[angle=0,width=9.0cm]{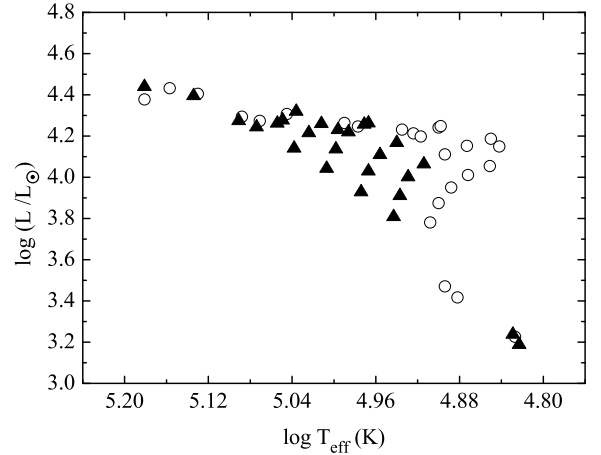}
\caption{Similar to Fig.~6, but for $M_{\rm WD,i}=1.0~M_\odot$.
The open circles, solid triangles denote systems with
$1.40~M_\odot\leq M_{\rm SN}<1.45~M_\odot$, and $M_{\rm
SN}\geq1.45~M_\odot$, respectively. } \label{FigVibStab}
\end{figure}

\begin{acknowledgements}
We are grateful to the anonymous referee for his/her constructive
suggestion improving this manuscript. This work was supported by
the Natural Science Foundation of China (NSFC) under grant number
10873011, and Program for Science \& Technology Innovation Talents
in Universities of Henan Province, China. ZH thanks the support of
NSFC (Grant No. 10821061), the Chinese Academy of Sciences (Grant
No. KJCX2-YW-T24) and the 973 project (Grant No. 2007CB815406).
\end{acknowledgements}


\begin{thebibliography}{}
\bibitem[Alexander \& Ferguson (1994)]{alex94} Alexander, D. R., \& Ferguson, J. W. 1994, ApJ, 437, 879
\bibitem[Astier et al. 2006]{asti06} Astier, P., et al. 2006, A\&A, 447, 31
\bibitem[Aubourg et al. 2008]{aubo08} Aubourg, E., Tojeiro, R., Jimenez, R., Heavens, A. F., Strauss, M.
A.,\& Spergel, D. N. 2008, A\&A, 492, 631
\bibitem[Chen \& Li 2007]{chen07} Chen, W.-C., \& Li, X.-D. 2007, ApJ, 658, L51
\bibitem[Chen \& Li 2009]{chen09} Chen, W.-C., \& Li, X.-D. 2009, \apj, 702, 686
\bibitem[Dewi et al. 2002]{dewi02} Dewi, J. D. M., Pols O. R., Savonije, G. J., \& van den Heuvel, E. P. J. 2002, MNRAS, 331, 1027
\bibitem[Di Stefano \& Kong (2003)]{di03} Di Stefano, R., \& Kong, A. K. H. 2003, ApJ, 592, 884
\bibitem[Dom\'{i}nguez et al. (2006)]{Domi06} Dom\'{i}nguez, I., Piersanti, L., Bravo, E., Tornamb\'{e}, A., Straniero, O., \& Gagliardi, S. 2006, ApJ, 644, 21
\bibitem[Dom\'{i}nguez et al. 1996]{Domi96} Dom\'{i}nguez, I., Straniero, O., Tornamb\`{e}, A., \& Isern, J. 1996, ApJ, 472, 783
\bibitem[Durisen 1977]{duri77} Durisen, R. H. 1977, \apj, 213, 145
\bibitem[Eggleton 1971]{egg71} Eggleton, P. P. 1971, MNRAS, 151, 351
\bibitem[Eggleton 1972]{egg72} Eggleton, P. P. 1972, MNRAS, 156, 361
\bibitem[Eggleton 1973]{egg73} Eggleton, P. P. 1973, MNRAS, 163, 279
\bibitem[Hachisu et al. 1996]{hach96} Hachisu, I., Kato, M., \& Nomoto, K. 1996, ApJ, 470, L97
\bibitem[Hachisu et al. 1999a]{hach99a} Hachisu, I., Kato, M., Nomoto, K., \& Umeda, H. 1999a, ApJ, 519, 314
\bibitem[Hachisu et al. 1999b]{hach99b} Hachisu, I., Kato, M., \& Nomoto, K. 1999b, ApJ, 522, 487
\bibitem[Hamuy et al. 1995]{hamu95} Hamuy, M., Phillips, M. M., Maza, J., Suntzeff, N. B., Schommer, R. A., \& Aviles, R. 1995, AJ, 109, 1
\bibitem[Hamuy et al. 1996]{hamu96} Hamuy, M., Phillips, M. M., Suntzeff, N. B., Schommer, R. A., Maza, J., \& Aviles, R. 1996, AJ, 112, 2391
\bibitem[Han et al. 1994]{han94} Han, Z., Podsiadlowski, P., \& Eggleton, P. P. 1994, MNRAS, 270, 121
\bibitem[Han \& Podsiadlowski 2004]{han04} Han, Z., \& Podsiadlowski, P. 2004, MNRAS, 350, 1301
\bibitem[Han \& Podsiadlowski 2006]{han06} Han, Z., \& Podsiadlowski, Ph. 2006, MNRAS, 368, 1095
\bibitem[Han 2008]{han08} Han, Z. 2008, ApJ, 677, L109
\bibitem[Hicken et al. 2007]{hick07} Hicken, M., et al. 2007, ApJ, 669, L17
\bibitem[Hillebrandt \& Niemeyer 2000]{hill00} Hillebrandt, W., \& Niemeyer, J, C. 2000, ARA\&A, 38, 191
\bibitem[Hillebrandt  et al. 2007]{hill07} Hillebrandt, W., Sim, S. A., \& R\"{o}pke, F. K. 2007, A\&A, 465, L17
\bibitem[Howell 2001]{how01} Howell, D. A. 2001, ApJ, 554, L193
\bibitem[Howell et al. 2006]{how06} Howell, D. A., et al., 2006, Nature, 443, 308
\bibitem[Hurley et al. 2000]{hurl00} Hurley, J. R., Pols, O. R., \& Tout, C. A. 2000, MNRAS, 315, 543
\bibitem[Hurley et al. 2002]{hurl02} Hurley, J. R., Tout, C. A., \& Pols, O. R. 2002, MNRAS, 329, 897
\bibitem[Iben \& Tutukov 1984]{iben84} Iben, I. Jr., \& Tutukov, A. V. 1984, ApJS, 54, 335
\bibitem[Iben \& Tutukov 1987]{iben87a} Iben, I. Jr.,  \& Tutukov, A. V. 1987, ApJ, 313, 727
\bibitem[Iben et al. 1987]{iben87b} Iben, I. Jr.,  Nomoto, K., Tornamb\`{e}, A., \& Tutukov, A. V. 1987, ApJ, 317, 717
\bibitem[Iben \& Tutukov 1991]{iben91} Iben, I. Jr.,  \& Tutukov, A. V. 1991, ApJ, 370, 615
\bibitem[Iben \& Tutukov 1994]{iben94} Iben, I. Jr.,  \& Tutukov, A. V. 1994, ApJ, 431, 264
\bibitem[Kahabka \& van den Heuvel 1997]{kaha97} Kahabka, P., \& van den Heuvel, E. P. J., 1997, ARA\&A, 35, 69
\bibitem[Kato \& Hachisu 1994]{kato94} Kato, M., \& Hachisu, I. 1994, ApJ, 437, 802
\bibitem[Kato \& Hachisu (2004)]{kato04} Kato, M., \& Hachisu, I. 2004, \apj, 613, L129
\bibitem[Kato et al. 2008]{kato08} Kato, M., Hachisu I., Kiyota S., \& Saio H. 2008, ApJ, 684, 1366
\bibitem[Langer et al. 2000]{lang00} Langer, N., Deutschmann, A., Wellstein, S., \& H\"{o}flich, P. 2000, A\&A, 362, 1046
\bibitem[Li \& van den Heuvel 1997]{li97} Li, X. -D., van den Heuvel, E. P. J. 1997, A\&A, 322, L9
\bibitem[Limongi \& Tornambe (1991)]{limo91} Limongi, M., Tornambe, A. 1991, ApJ, 371, 317
\bibitem[L\"{u} et al. 2009]{lu09} L\"{u}, G., Zhu, C., Wang, Z., \& Wang, N. 2009, MNRAS, 396, 1086
\bibitem[Meng et al. (2008)]{meng08} Meng, X., Chen, X., \& Han, Z. 2008, A\&A, 487, 625
\bibitem[Meng et al. 2009]{meng09} Meng, X., Chen, X., \& Han, Z. 2009, MNRAS, 395, 2103
\bibitem[Meng \& Yang 2010a]{meng10a} Meng, X., \& Yang, W. 2010, MNRAS, 401, 1118
\bibitem[Meng \& Yang 2010b]{meng10b} Meng, X., \& Yang, W. 2010, ApJ, 710, 1310
\bibitem[Mereghetti et al. 2009]{mer09} Mereghetti, S., Tiengo, A., Esposito, P., et al. 2009, Science, 325, 1222
\bibitem[Narayan \& Pophm 1989]{nara89} Narayan, R., \& Popham, R. 1989, ApJ, 346, L25
\bibitem[Nomoto 1982]{nom82} Nomoto, K. 1982, \apj, 253, 798
\bibitem[Nomoto et al. 1997]{nom97} Nomoto, K., Iwamoto, K., \& Kishimoto, N. 1997, Science, 276, 1378
\bibitem[Perlmutter et al. 1999]{perl99} Perlmutter, S., et al., 1999, \apj, 517, 565
\bibitem[Piersanti et al. 2003]{pier03} Piersanti, L., Gagliardi, S., Iben, I. Jr., \& Tornambe, A. 2003,
ApJ, 598, 1229
\bibitem[Podsiadlowski et al. 2008]{pod08} Podsiadlowski, P., Mazzali, P., Lesaffre, P., Han, Z., \& Forster, F. 2008, NewAR, 52, 381
\bibitem[Pols et al. 1995]{pol95} Pols, O. R., Tout, C. A., Eggleton, P. P., \& Han, Z. 1995, MNRAS, 274, 964
\bibitem[Pols et al. 1998]{pol98} Pols, O. R., Schroder, K. P., Hurly, J. R., \& Tout, C. A. 1998, MNRAS, 298, 525
\bibitem[Riess et al. 1998]{ries98} Riess, A. et al., 1998, AJ, 116, 1009
\bibitem[Ritter 1985]{ritt85} Ritter, H. 1985, A\&A, 148, 207
\bibitem[Rogers \& Iglesias (1992)]{roge92} Rogers, F. J., \& Iglesias, C. A. 1992, ApJS, 79, 507
\bibitem[R\"{o}pke \& Hillebrandt 2005]{rop05} R\"{o}pke, F. K., \& Hillebrandt, W. 2005, A\&A, 431, 635
\bibitem[Scalzo et al. 2010]{scal10} Scalzo, R. A., et al. 2010, ApJ, 713, 1073
\bibitem[Silverman et al. 2010]{silv10} Silverman, J. M., et al. 2010, MNRAS, submitted [arXiv:1003.2417]
\bibitem[Tanaka et al. 2010]{tana10} Tanaka, M., et al. 2010, ApJ, 714, 1209
\bibitem[Tornamb\`{e} \& Matteucci 1986]{torn86} Tornamb\`{e}, A., Matteucci, F. 1986, MNRAS, 223, 69
\bibitem[Tutukov \& Yungelson 1994]{tutu94} Tutukov, A. V., \& Yungelson, L. R. 1994, MNRAS, 268, 871
\bibitem[Wang et al. 2008]{wang08} Wang, B., Meng, X., Wang, X., \& Han, Z. 2008, ChJAA, 8, 71
\bibitem[Wang et al. (2009a)]{wang09a} Wang, B., Meng, X., Chen, X., \& Han, Z. 2009a, MNRAS, 395, 847
\bibitem[Wang et al. 2009b]{wang09b} Wang, B., Chen, X., Meng, X., \& Han, Z. 2009b, ApJ, 701, 1540
\bibitem[Wang \& Han 2009]{wang09c} Wang, B., \& Han, Z. 2009, A\&A, 508, L27
\bibitem[Wang et al. 2010a]{wang10a} Wang, B., Li, X.-D., \& Han, Z. 2010a, MNRAS, 401, 2729
\bibitem[Wang \& Han (2010b)]{wang10b} Wang, B., \& Han, Z. 2010b, A\&A, 515, A88
\bibitem[Wang \& Han (2010c)]{wang10c} Wang, B., \& Han, Z. 2010c, RAA, 10, 681
\bibitem[Webbink 1984]{webb84} Webbink, R. F. 1984, ApJ, 277, 355
\bibitem[Whelan \& Iban 1973]{whel73} Whelan, J., \& Iban, I. Jr. 1973, ApJ, 186, 1007
\bibitem[Woosley et al. 1986]{woos86} Woosley, S. E., Taam, R. E., \& Weaver, T. A. 1986, \apj, 301, 601
\bibitem[Woudt et al. 2009]{wou09} Woudt, P. A., Steeghs, D., Karovska, M., et al. 2009, \apj, 706, 738
\bibitem[Yoon \& Langer 2003]{yoon03} Yoon, S. -C., \& Langer, N. 2003, A\&A, 412, L53
\bibitem[Yoon et al. 2004]{yoon04a} Yoon, S. -C., Langer, N., \& Scheithauer, S. 2004, A\&A, 425, 217
\bibitem[Yoon \& Langer 2004]{yoon04b} Yoon, S. -C., \& Langer, N. 2004, A\&A, 419, 623
\bibitem[Yoon \& Langer 2005]{yoon05} Yoon, S. -C., \& Langer, N. 2005, A\&A, 435, 967
\bibitem[Yuan et al. 2007]{yuan07} Yuan, R. F., et al. 2007, Astron. Telegr., 1212



\end{thebibliography}
\end{document}